# Device and method for investigation of mechanical properties of fibers under high-strain rate tensile load.

**Sergey Lopatnikov, Nikolas Shevchenko, John W. Gillespie Jr**

**Abstract**. A new apparatus and method is proposed for the investigation of material behavior under high-strain-rate tensile loads. We refer this apparatus as a Split Flying Bar. The method is based on using the inertia of a working mass attached to a specimen. The specimen is placed between the working mass (front part of the bar) and backing part of the bar, which is captured in flight by special brake. When the back part of the flying split bar is stopped, the working mass continues the flight by inertia, creating specimen tension with strain rate depending on the length of specimen and velocity of flying bar. Properly choosing the working mass and the speed of the flying bar, one can tune maximal stress and strain rate over a wide range. The method is highly scalable and can be used for investigation of specimens from single filament up to reasonable macroscopic size. Contrary to tensile SHPB, the method provides a way to investigate materials with practically arbitrary strain to failure. For example, polyuria, whose strain to failure reaches hundreds of percent. Also, the method can be used for high-strain rate pull-out tests or to measure the quality of adhesion layers under high-stress rates, e.t.c. In this paper I introduce the basics of the method and interpretation of the data.

**Introduction.**

The ability to characterize mechanical behavior of the materials under high strain rates, both under compression and tension, is crucial for the material sciences. It not only helps to predict the behavior of the materials under different loading conditions, but also it leads to the deeper understanding of physical and chemical mechanisms underlying the material performance. In turn, such understanding plays the key role in improvement of existing materials and creation of the new ones with required properties.

Since the late 40's, Split Hopkinson Pressure or Tensile Bars (Davies R.M. 1948; Kolsky, 1949; Lindholm & Yekley, 1964; Gray III, 2000; Lopatnikov et al., 2004 a)) have become the most common methods of investigation of material behavior under high-strain rates. The creation of compression stress and associated data reduction are simple and has solid physical background (Gray III, 2000; Lopatnikov et al., 2004 b , Lopatnikov et al, 2011). However, tensile split bars are significantly more complicated devices and reduction of their data causes serious questions. In the meantime, other mechanical tensile devices cannot provide high enough strain rate loads.

The Split Flying Bar method is based on the use of inertial forces to create the necessary stress level and strain rate. The principal construction of the Split Flying Bar is shown in Figure 1.

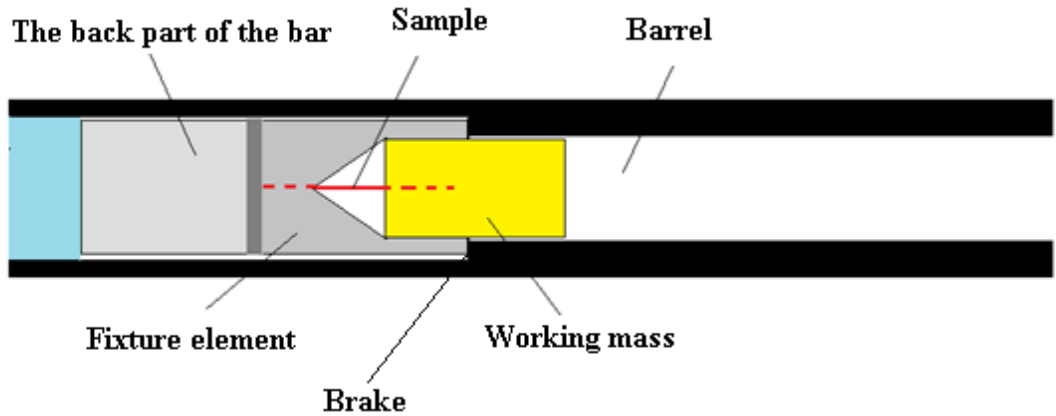

**Figure 1.** *Principal construction of the Split Flying Bar for high-strain rate tensile experiments.*

The principle is simple: The split bar, with the specimen fixed between working mass and back part of the bar, is accelerated by a device such as a (gas) gun, rocket, spring, gravity, etc. After reaching the necessary speed, the backing part having larger diameter is stopped by the brake. In the simplest case the brake can be the taper to the narrower part of the barrel. The working mass, continuing to move forward by inertia, leads to the tensile stress of the sample. The strain rate is dependent on the speed of the flying bar and value of working mass.

One can choose the kinetic energy of the working mass in accordance with the energy required to break the sample. If the kinetic energy of the working mass is chosen so that:

$$\frac{M_w \cdot V^2}{2} \gg U_{cr} \qquad (1)$$

Where $U_{cr}$ is the energy of sample destruction. It guaranties that the sample will be destroyed and that the effective strain rate $\dot{\varepsilon}_{eff}$ will be practically constant during the deformation of the sample.

$$\dot{\varepsilon}_{eff} \sim \frac{V}{L} \qquad (2)$$

Oppositely, if kinetic energy of the working mass is chosen so that:

$$\frac{M_w \cdot V^2}{2} < U_{cr} \qquad (3)$$

One will observe both loading and unloading behavior of the specimen. Under this condition the sample will not be destroyed. Changing the working mass simultaneously with the speed of the Flying Bar so that kinetic energy remains the same, one can obtain the set of equi-energy stress-strain diagrams completely representing the rheology of the sample material. It is important in these experiments that the strain rates can vary under the same energy of impact in the wide range. For example, if the length of specimen is equal to 1 cm and velocity of the bar is equal to only 10 m/s, the strain rate will be equal to $10^3$. In the meantime the Flying Bar can easily reach speeds of 20, 50, 100 m/s or even higher, which would result in strain-rates in the range of $10^4$ and higher.

It is clear that in comparison with the Hopkinson bar, the proposed method permits investigation of high strain-rate rheological properties of the materials with arbitrary strain to failure. In experiments with the Hopkinson bar this value is limited by the speed of the striker bar and its length (defining the duration of pulse). In the Flying Bar method, the working mass can reach any distance, which is necessary to destroy the sample (or to reach necessary level of sample elongation). Addition flexibility is related with the fact that dependent on the specific task, one could measure various physical quantities using different methods of registration. For example, one could directly measure the elongation of the sample by optical means, or alternatively measure displacement or velocity of the working mass. Likewise, one could measure the deceleration of working mass and then calculate the associated force, or use a load cell, etc.

Let us consider briefly the physical background of the Split Flying Bar data reduction. First, similarly to SHPB experiments, there is an obvious requirement for experiment. The sample being investigated must be in approximate mechanical equilibrium. This means than non-homogeneity of the stress distribution within the specimen must be negligibly small in comparison with (homogeneous) average stress.

Since the pioneering publication by Kolsky (1949), the common view is that with sufficient accuracy a specimen comes to mechanical equilibrium after 3 to 4 times the period for sound to go through the specimen. A better physically grounded consideration (Lopatnikov et.al, ) shows, however, that one has to establish limits on the strain rate, rather than to wait enough time. Specifically, the strain rate during the whole experiment must satisfy the condition:

$$\dot{\varepsilon} << \dot{\varepsilon}_{max} = \frac{C}{L} \qquad (4)$$

$C$ denotes the "process speed", which is not necessarily just the speed of sound. It can be the speed of any (the slowest) wave-type process responsible for sample elongation (Lopatnikov et al. ...).

Considering established equilibrium, if one can neglect the mass of sample in comparison with the working mass, then the suggested experimental scheme can be described by the simple "pendulum" model: in general one can consider a specimen as a non-linear spring:

$$M_w \frac{d^2 X_w(t)}{dt^2} = F(X_w(t)) \equiv -S\sigma(X_w) \qquad (5)$$

Here - $X_W$ - is the coordinate of the working mass.

In accordance with (5), to get the loading curve of the specimen, it is enough to measure only one value: the speed of the working mass as a function of time. In this case, one obtains the strain of the sample as:

$$\varepsilon(t) = \frac{X_w(t)}{L} = \frac{1}{L}\int_0^t V(t)\,dt \tag{6}$$

And stretching stress simply by one-time differentiation of speed:

$$\sigma(t) = -\frac{M}{S}\frac{dV(t)}{dt} \tag{7}$$

The speed of the working mass can be measured with very high accuracy by many methods. Some possibilities are optical "wind shadowing", Electromagnetic induction using a pickup coil if the working mass is made from magnetic material (or contains magnetic inclusions) or laser Doppler velocimetry and interferometry, etc.

For the sake of increasing method accuracy, one can use independent methods for measuring the sample elongation and developed stress. The optimal set of the measurements depends on the specific test.

According to these suggestions, the energy, which is necessary to destroy the fiber, is equal to:

$$E = \frac{1}{2} S \cdot L \cdot K \varepsilon_{max}^2 \tag{8}$$

Considering the initial strain rate $\dot{\varepsilon} = \omega \equiv \frac{V}{L}$, the working mass, which is necessary to get appropriate kinetic energy is equal to:

$$M = \frac{2E}{\omega^2 L^2} = \frac{S \cdot L \cdot K \varepsilon_{max}^2}{\omega^2 \cdot L} \tag{9}$$

In Figure 2 we present a model calculation of the working mass needed to break the sample as a function of the strain rate. The plots are for a 20 micron diameter fiber, having 1 cm-length, strength equal to 186 GPa (which is typical for Kevlar129) and 3% strain to failure, and for polyuria with strength 15.2 MPa and strain to failure 300%, suggesting linear behavior of the sample up to the failure.

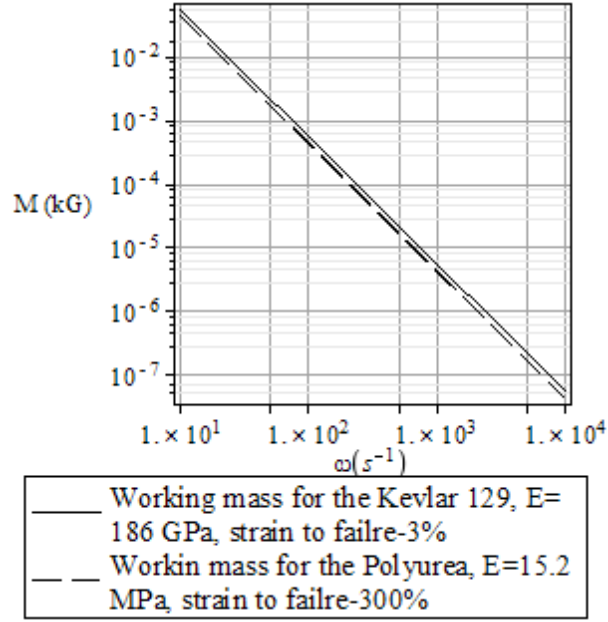

**Figure 2.** *Working mass necessary to break the Kevlar-129 and Polyurea 20-micron fiber of 1 cm length.*

One can see that both masses are close to each other. Thus, to reach a strain-rate of $10^3 \, s^{-1}$ and break the 20 micron diameter 1 cm long fiber one needs a mass approximately equal to 0.05 g. If, for example, the working mass has a 2 mm diameter and is made from aluminum, then its length must be equal to approximately 6 mm. This shows that the fiber-testing set-up can be made pretty compact and simple. There are no significant problems to reach strain-rates in the range of $10^4 \, s^{-1}$, however, higher strain rates will need to account for the mass of the specimen itself.

Let us briefly consider the dynamics of the working mass. Equation (5) can be solved for practically any stress-strain-strain-rate diagram. Using Tichonov's methods of data reduction materials with memory can also be studied. We consider the most trivial case in which the sample behaves linearly. In this case, the system becomes a simple linear oscillator with the frequency $\Omega = \sqrt{\dfrac{SE}{M_w}}$ and has the obvious "pendulum" solution: $V(t) = V_0 \cos \Omega t$. If the working mass is chosen in accordance with (9), the specimen breaks down exactly when the speed of the specimen $V(t)$ becomes equal to zero. Strain of the sample during the experiment is equal in this case to $\varepsilon(t) \approx \dfrac{V_0}{\Omega} \sin \Omega t$, strain rate to $\dot{\varepsilon}(t) \approx \dfrac{V_0}{L} \sin \Omega t$ and stress $\sigma(t) \approx \Omega V_0 M_w \sin \Omega t$.

It is useful to mention that by excluding time from these relationships one can obtain stress-strain-strain-rate diagrams $\sigma = \sigma(\varepsilon, \dot{\varepsilon})$, which we proposed to use earlier for representation of the rheological data of the material with complex behavior such as Shear Thickening Fluid (STF) (Lim, ).

If one chooses the working mass significantly bigger than followed from (9), the speed of working mass, and thus the strain-rate will be practically constant up to the destruction of specimen. If one chooses this mass smaller than the mass defined by (9), the specimen will not be destroyed and oscillations of working mass will be observed. One will be able to measure up-load-down-load characteristics of the material, physical energy absorption, progressive failure, etc.

It is clear that proposed method of SLFB can not only be used for obtaining the stress-strain-strain-rate diagrams of the materials, but for many other experiments. For example, to prove high-strain-rate pull-out experiments, investigate high-strain-rate behavior of adhesion layers, connecting fibers and matrix and effect of nano-inclusions.

Method can also be used for macro-specimens and under ballistic speeds.

## Conclusion.

We propose a method for investigation of material behavior under high strain-rate tension. The method is highly scalable and can be applied to a wide range of problems.

## Literature.